\documentclass[11pt]{article}
\begin{document}

\centerline{Evaluation of Spectra of Baryons Containing Two Heavy
Quarks in Bag Model}

\vspace{0.8cm}

\centerline{\small Da-Heng He$^1$, Ke Qian$^1$, Yi-Bing
Ding$^{2,6}$, Xue-Qian Li$^{1,5,6}$ and Peng-Nian Shen$^{4,3,5,6}$
}

\vspace{0.4cm}

\noindent 1. Department of Physics, Nankai University, Tianjin
300071, China;\\

\noindent 2. Graduate School of The Chinese Academy of Sciences,
Beijing, 100039, China,\\

\noindent 3. Institute of High Energy Physics, CAS, P.O. Box
918(4), Beijing 100039, China\\

\noindent 4. Center of Theoretical Nuclear Physics, National
Laboratory of Heavy Ion Accelerator, Lanzhou 730000, China\\

\noindent 5. Institute of Theoretical Physics, CAS, P.O. Box 2735,
Beijing, 100080, China.\\

\noindent 6. China Center of Advanced Science and Technology
 (World Laboratory), P.O.Box 8730, Beijing 100080, China

\vspace{1cm}

\begin{abstract}
In this work, we evaluate the energy spectra of baryons which
consist of two heavy and one light quarks in the MIT bag model.
The two heavy quarks constitute a heavy scalar or axial vector
diquark. Concretely, we calculate the spectra of $|q(QQ')>_{1/2}$
and $|q(QQ')>_{3/2}$ where $Q$ and $Q'$ stand for $b$ and/or $c$
quarks. Especially, for $|q(bc)>_{1/2}$ there can be a mixing
between $|q(bc)_0>_{1/2}$ and $|q(bc)_1>_{1/2}$ where the
subscripts 0 and 1 refer to the spin state of the diquark (bc),
the mixing is not calculable in the framework of quantum mechanics
(QM) as the potential model is employed, but can be evaluated by
the quantum field theory (QFT). Our numerical results indicate
that the mixing is sieable
\end{abstract}

\vspace{1cm}

\section{ Introduction}

At present, the non-perturbative QCD which dominates the low
energy physics phenomena, is not fully understood yet, a
systematic and reliable way for evaluating the non-perturbative
QCD effects on such as the hadron spectra and hadronic matrix
elements is lacking. Fortunately, however, for the heavy flavor
mesons or baryons which at least contain one b or c quarks
(antiquarks) the situation becomes simpler due to an extra
$SU_f(2)\otimes SU_s(2)$ symmetry\cite{Isgur}. The studies in this
field provide us with valuable information about the QCD
interaction and its low energy behavior.

Among all the interesting subjects, the hadron spectra would be
the first focus of attention. The spectra of the $J/\psi$ and
$\Upsilon$ families have been thoroughly investigated in different
theoretical approaches. Commonly, the spectra are evaluated in the
potential model inspired by QCD, where the QCD Coulomb-type
potential is directly derived from the one-gluon-exchange
mechanism, and the confinement term originating from the
non-perturbative QCD must be introduced by hand\cite{Rosner}. For
the heavy quarkonia where only heavy quark flavors are involved,
the potential model definitely sets a good theoretical framework
for describing such systems where relativistic effects are small
compared to the mass scale. An alternative model, the bag model
can also provide a reasonable confinement for quarks. In fact, for
light hadrons, especially light baryons, the bag model may be a
better framework for describing their static behaviors. The MIT
bag model has some advantages \cite{Jaffe}. First, even though it
does not hold a translational invariance, the quarks inside the
hadron bag obey the relativistic Dirac equation and moreover, they
can be described in the Quantum Field Theory (QFT) framework,
namely there exist creation and annihilation operators for the
constituents of the hadron. The latter property is very important
for this work, that we can calculate a mixing between
$|q(bc)_0>_{1/2}$ and $|q(bc)_1>_{1/2}$ (the notations will be
explained below) states and it is impossible in the Quantum
Mechanics (QM) framework.

Another interesting subject is if the diquark structure which
consists of two quarks and resides in a color-anti-triplet $\bar
3$, exists in baryons. Its existence, in fact, is till in dispute.
For light diquark which is composed of two light quarks, the
relativistic effects are serious and the bound state should be
loose. By contraries, two heavy quarks (b and c) can constitute a
stable bound state of $\bar 3$, namely a diquark which serves as a
source of static color field\cite{Falk}. As a matter of fact, the
un-penetrable bag boundary which provides the confinement
conditions to the constituents of the hadron, is due to the
long-distance non-perturbative QCD effects, to evaluate the
spectra, one needs to include the short-distance interaction
between the constituents and it can be calculated in the framework
of the perturbative QCD. In this work, we are going to evaluate
the spectra of baryons which contains two heavy quarks (b and/or
c) and a light quark and take the light-quark-heavy-diqaurk
picture which obviously is reasonable for the case of concern.

For evaluating the hadron spectra, the traditional method is the
potential model. For baryons, the quark-diquark picture can reduce
the three-body problem into a two-body problem and leads to a
normal Schr\"odinger equation. Solving the Schr\"odinger equation,
one can get the binding energy of quark and
diquark\cite{Ebert,Tong}. In recent years, remarkable progresses
have been made along this direction. The authors of refs.
\cite{Kiselev} have carefully studied the short-distance and
long-distance effects, then derived a modified potential and
obtained the spectroscopy of the baryons which contain two heavy
quarks by using the non-relastivistic Schr\"odinger equation.
Meantime, in Ebert et al.'s new work, the light quark is treated
as a fully relativistic object and the potential is somehow
different from that in their earlier work\cite{Ebert2}. In their
works, not only the ground states of such baryons are obtained,
but also the excited states are evaluated.

However, the potential model has two obvious drawbacks. First,
even though the diquark is heavy, the constituent quark mass of
the light quark is still comparable to the linear momentum which
is of order of $\Lambda_{QCD}$. Thus the reduced mass is not large
and the relativistic effects are still significant. Secondly,
working in the framework of QM, it is impossible to estimate the
mixing of $|q(bc)_0>_{1/2}$ and $|q(bc)_1>_{1/2}$ where the
subscripts 0 and 1 of the (bc)-diquark denote the total spin of
the subsystem, i.e. the (bc)-diquark (we only consider the ground
state of $l=0$). The reason is that there are no creation and
annihilation operators in the traditional QM framework, so the
transition $(bc)_1+q\rightarrow (bc)_0+q$, i.e. $A+q\rightarrow
S+q$ where the notation A and S refer to the axial-vector and
scalar diquarks respectively, is forbidden, even though the
transition is calculable in QFT.

On other side, the MIT bag model does not suffer from the two
drawbacks. In this picture, since the diquark is heavy, it hardly
moves so that can be supposed to sit at the center of the bag,
whereas the light quark is freely moving in the bag and its
equation of motion is the relativistic Dirac equation with a
certain boundary condition\cite{Jaffe} and both the quark and
diquark are quantized in the QFT. Thus the relativistic effects
are automatically included. Secondly, one can deal with a possible
conversion of the constituents in the bag in terms of QFT, namely
one can let a constituent be created or annihilated, thus the
transition $A+q\rightarrow S+q$ is allowed and the corresponding
mixing of $|q(bc)_0>_{1/2}$ and $|q(bc)_1>_{1/2}$ is calculable.

Usually the bag model is not very applicable to the light mesons
because the spherical boundary is not a good approximation for the
two-body system. Even though the quark-diquark structure is a
two-body system, the aforementioned problem does not exist because
the diquark is much heavier than the light quark. The picture is
in analog to the solar system or an atom where only one valence
electron around the heavy nucleus, and  spherical boundary would
be a reasonable choice.

In this work, following the literature \cite{Jaffe}, we treat the
short-distance QCD interaction between the light quark and heavy
diquark  perturbatively. Since the interaction energy $E_{int}(R)$
is not diagonal for $|q(bc)_1>_{1/2}$ and $|q(bc)_0>_{1/2}$, we
may diagonalize the matrix to obtain the eigenvalues and
eigenfunctions which would be the masses of the baryons with
flavor $q(bc)$ and spin 1/2. Moreover, for the other baryons
$|q(bb)_1>_{1/2(3/2)}$ $|q(cc)_1>_{1/2(3/2)}$ $|q(bc)_1>_{3/2}$,
the diquark must be an axial vector due to the Pauli
principle\cite{Close}.

The paper is organized as follows, after the introduction, we
derive all the formulation of $E_{int}(R)$ and $M_B$ in Sec.II,
then in Sec.III we present the numerical results and all concerned
parameters, finally the last section is devoted to discussions.\\

\section{  Formulation}

\noindent 1. A brief review of the MIT bag model

The wavefunction of a light quark in the MIT bag  obeys the Dirac
equation for free fermion and a boundary condition which forbids
the quark current to penetrate the bag boundary. It has a
form\cite{Jaffe}
\begin{equation}
q(\bf{r},t)=\frac{N(\chi)}{\sqrt{4\pi}} \left (
\begin{array}{cc}
(\frac{\omega+m}{\omega})^{1/2}ij_{0}(\chi r/R)U
\\-(\frac{\omega-m}{\omega})^{1/2}j_{1}(\chi r/R){\mbox{\boldmath $\sigma$}}\cdot {\bf r}U,
\end{array}
\right )
\end{equation}
with
\begin{equation}
N^{-2}(\chi)=R^{3}j_{0}^{2}(\chi)\frac{2\omega(\omega-1/R)+m/R}{\omega(\omega-m)},
\end{equation}
where $j_l$ is a spherical Bessel function, $U$ is a two-component
Pauli spinor, the eigen-energy is
\begin{equation}
\omega(m,R)=\frac{[\chi^{2}+(mR)^{2}]^{1/2}}{R},
\end{equation}
and the eigenvalue $\chi$ satisfies an equation
\begin{equation}
\tan(\chi)=\frac{\chi}{1-mR-[\chi^{2}+(mR)^{2}]^{1/2}}.
\end{equation}

In our picture, the two heavy quarks constitute a diquark which is
a boson-like bound state of color $\bar 3$ and because it is
heavy, it hardly moves, but sits at the center of the bag. Its
wavefunction can be written as\cite{Ebert1}
\begin{eqnarray}\label{heavy}
\psi(r) &=& {N\over 4\pi}e^{{\Lambda r^2\over 2}}\;\;\;\;\;\; {\rm
for\;the\;scalar\;diquark};\\
\psi_{\mu}(r) &=& {N\over 4\pi}e^{{\Lambda r^2\over
2}}\eta_{\mu}\;\;\;\;\;\; {\rm for\;the\;axial\; vector\;diquark},
\end{eqnarray}
where $\eta_{\mu}$ is the polarization vector which is normalized
as $\eta^2=-1$, $\Lambda$ is a parameter and will be discussed
later in the text.\\

\noindent 2. Formulation for the baryon spectra.

In the CM frame of the baryon, the total mass of the baryon can be
written as
\begin{equation}
M_{B}=M_{D}+\omega+E_{int}(R)+{4\over 3}\pi R^{3}B-{z\over R}
\end{equation}
where $E_{int}(R)$ is the interaction energy between the diquark
and light quark, $B$ is the bag constant and $z$ is a constant for
the zero-point energy. $\omega$ and $M_D$ are the eigen-energy of
the free light quark and the mass of the heavy diquark which are
given in the literature\cite{Jaffe,Ebert}. In this work, following
the standard procedure \cite{Landau,Jaffe}, we are going to
calculate the interaction energy $E_{int}(R)$. Generally, the
interaction hamiltonian in the bag can be expressed as
\begin{equation}\label{HDD}
H_{D'D}=-\frac{\lambda_{1}^{a}}{2}\frac{\lambda^{a}_{2}}{2}g_{s}^{2}
\int\overline{q}({\bf x})\gamma^{\mu}q({\bf
x})D_{\mu\nu}\Psi^{\ast}({\bf y})
\frac{<D'|J^{\nu}|D>}{\sqrt{MM'}}\Psi({\bf y})d^{3}xd^{3}y.
\end{equation}
It is noted that without mixing, i.e. $H_{D'D}$ is diagonal,
$E_{int}=H_{D'D}$, however, if there is non-diagonal $H_{D'D}$,
$E_{int}$ is the eigenvalues of the hamiltonian matrix. The
expectation value of the Casimir operator
$<0|\lambda_{1}^{a}\lambda_{2}^{a}|0>=-16/3$, the strong coupling
$g_s^{2}=4\pi\alpha_{s}$, $q({\bf x})$ is the wavefunction of the
free light quark, $<D'|J^{\nu}|D>$ is the effective vertex for
$DD'g$ and $D_{\mu\nu}$ is the gluon propagator in the Coulomb
gauge\cite{Lee}. The form of such a propagator in the
configuration space reads
\begin{eqnarray}
G(\mathbf{r},\mathbf{r}_{0})=\frac{1}{4\pi}[\frac{1}{|\mathbf{r}-
\mathbf{r}_{0}|}+\sum_{l=0}^{\infty}\
\frac{(l+1)(1-k)}{l+(l+1)k}\frac{(rr_{0})^{l}}{R^{2l+1}}P_{l}(cos\theta)].
\end{eqnarray}
In this work,  $\mathbf{r_{0}}$ is small, thus the main
contribution comes from the $l=0$ component. Then, the expression
can be simplified as:
\begin{eqnarray}
G(\mathbf{r},\mathbf{r}_{0})=\frac{1}{4\pi}[\frac{1}{|\mathbf{r}-
\mathbf{r}_{0}|}+\frac{\tilde{k}}{R}].
\end{eqnarray}
It is noted that the term $\frac{\tilde{k}}{R}$ is due to the
mirror charge effect\cite{Lee}, in fact, it indeed corresponds to
the zero-point energy in the bag as Jaffe et al.
suggested\cite{Jaffe}. Actually, $\tilde{k}$ is related to the
vacuum property and still serves as a free parameter that cannot
be determined from any underlying theory yet. We choose
$\tilde{k}=0.87$ to fit the most recent lattice result about the
spectrum of $(ccq)_{\frac{1}{2}^{+}}$.

\noindent 3. The D'Dg effective vertex

If we only consider the ground states of the diquark, namely the
two heavy quarks are in $l=0$ color-anti-triplet state, the
diquark can be either a scalar (denoted as $S$) with $s=0$ or an
axial vector (denoted as $A$) of $s=1$. The effective vertices can
be derived by the quantum field theory under the heavy quark
limit\cite{Guo}. The effective vertex for $SS'g$ is
\begin{equation}
<S'|J^{\nu}|S>=\sqrt{MM'}(f_{1}v'^{\nu}+f_{2}v^{\nu})
\end{equation}
and the $AA'g$ effective vertex is of the form
\begin{eqnarray}
<A'|J^{\nu}|A> &=& \sqrt{MM'}[f_{3}(\eta\cdot\eta'^{\ast})v'^{\nu}
+f_{4}(\eta'^{\ast}\cdot\eta)v^{\nu} +f_{5}(\eta\cdot
v')(\eta'^{\ast}\cdot v)v'^{\nu}
 \nonumber \\&& {}+f_{6}(\eta\cdot v')(\eta'^{\ast}\cdot v)v^{\nu}
+f_{7}{\eta'^{\ast}}^{\nu}(\eta\cdot v')+f_{8}(\eta'^{\ast}\cdot
v)\eta^{\nu}]
\end{eqnarray}
where $v,\; v',\; \eta_{\mu}$ and $\eta_{\mu}'$ are the
four-velocities and polarization vectors of $D$ and $D'$
respectively, $f_i'$s are the form factors at the vertices. As we
did in our previous work \cite{Tong} where the effective potential
model was employed, for convenience of calculation, we would write
the polarization into a spin-operator which acts on the
wavefunction of the axial-vector diquark as
\begin{equation}
\eta_{\mu}={1\over\sqrt {2}}({\mbox{\boldmath $\beta$}}\cdot {\bf
s},{\bf s}),
\end{equation}
where higher order relativistic corrections proportional to ${{\bf
p}^2\over M_D^2}$ are neglected and the factor $1/\sqrt 2$ is a
normalization factor because $<s^2>=s(s+1)=2$. It is worth
noticing that $s_i$ and $s_j$ do not commute with each other, thus
one must be careful about their order in deriving formula.

The effective vertex for ASg is written as
\begin{eqnarray}
<A'|J^{\nu}|S> &=&
  \sqrt{MM'}[f_{11}{\eta'^{\ast}}^{\nu}+f_{12}(\eta'^{\ast}\cdot
v)v'^{\nu}
 \nonumber \\&& {}+f_{13}(\eta'^{\ast}\cdot v)v^{\nu}
+f_{14}i\epsilon^{\nu l \rho
\sigma}\eta'^{\ast}_{l}v'_{\rho}v_{\sigma}].
\end{eqnarray}

Here we must stress that for convenience of calculation, we turn
$\eta_{\mu}$ into the quantum spin-operator as evaluating the
interacting energy for $|q(QQ')>$, but as pointed above,  we
cannot calculate the mixing between $|q(bc)_1>_{1/2}$ and
$|q(bc)_0>_{1/2}$ in QM, but need to carry out the derivation in
QFT instead. Then, we have to keep the polarization $\eta_{\mu}$
in a 4-vector form as
$$\eta_{\mu}^{\pm}=\mp {1\over\sqrt 2}(0,1,\pm i,0)\;\;{\rm
and}\;\; \eta^0_{\mu}=(0,0,0,1)$$. Because the diquark is very
heavy and hardly moves, $|{\mbox{\boldmath $\beta$}}|\ll 1$, one
can use the polarization vector in the reference frame where the
diquark is at rest.

In the heavy quark limit, we have
\begin{eqnarray}
 && f_{1}=f_{2}=f_{7}=f_{8}=-f_{3}=-f_{4}=f_{14}=1
 \nonumber \\&& {}f_{11}=f_{12}=f_{13}=0,
\end{eqnarray}
and for very heavy diquark the above approximation
holds\cite{Guo}.

4. The interaction energy

To calculate the interaction energy, the basic formula is
eq.(\ref{HDD}). For the heavy diquark, $|{\bf p}\ll M_D$, thus the
relativistic corrections proportional to and higher than ${{\bf
p}^2\over M_D^2}$ can be safely ignored.

Based on the approximations, one can easily obtain the interaction
energies.

For  the baryon where the diquark is a scalar of $\bar 3$, the
interaction energy between the light quark and the scalar diquark
is corresponding to the transition matrix element of
$<q,S|H_{eff}|q,S>$, and can be expressed as
\begin{eqnarray}
H_{SS} &=&
-\frac{\lambda_{1}^{a}}{2}\frac{\lambda_{2}^{a}}{2}\int\int\overline{q}(\mathbf{x})\gamma^{\mu}q(\mathbf{x})D_{\mu\nu}\Psi^{\ast}(\mathbf{y})
[f_{1}v'^{\nu} +f_{2}v^{\nu}]\Psi(\mathbf{y})d^{3}xd^{3}y\nonumber \\
{}&=&g^{2}_{s}N\int^{R}_{0}[j_{0}^{2}(\frac{\chi
r_{x}}{R})+j_{1}^{2}(\frac{\chi r_{x}}{R})]\frac{2}{r_{x}}d^{3}x,
\end{eqnarray}
where ${\bf x}$ is the spatial coordinate of the light quark,
${\bf y}$ is that of the heavy diquark, ${\bf r}={\bf x}-{\bf y}$
is the relative coordinate of the quark and diquark.

For the baryon where the diquark is an axial vector, the matrix
element  $<q,A|H_{eff}|q,A>$ is written as formula of
$H_{AA\frac{1}{2}}$:
\begin{eqnarray}
H_{AA\frac{1}{2}} &=&
-\frac{\lambda_{1}^{a}}{2}\frac{\lambda_{2}^{a}}{2}\int\int\overline{q}(\mathbf{x})\gamma^{\mu}q(\mathbf{x})D_{\mu\nu}\Psi^{\ast}(\mathbf{y})
[f_{3}(\eta\cdot\eta'^{\ast})v'^{\nu}
+f_{4}(\eta'^{\ast}\cdot\eta)v^{\nu} \nonumber \\&&
{}+f_{5}(\eta\cdot v')(\eta'^{\ast}\cdot v)v'^{\nu}
+f_{6}(\eta\cdot v')(\eta'^{\ast}\cdot v)v^{\nu}
+f_{7}\eta'^{\ast}(\eta\cdot v')\nonumber \\&&
{}+f_{8}(\eta'^{\ast}\cdot
v)\eta^{\nu}]\Psi(\mathbf{y})d^{3}xd^{3}y\nonumber \\
{}&=&g^{2}_{s}N\int^{R}_{0}[j_{0}^{2}(\frac{\chi
r_{x}}{R})+j_{1}^{2}(\frac{\chi
r_{x}}{R})]d^{3}x\int(-\frac{1}{|\mathbf{r_{x}-\mathbf{r_{y}}}|}+\frac{k}{R})|\Psi(y)|^{2}d^{3}y
\nonumber
\\&& {}+g^{2}_{s}\frac{C_{\alpha\beta}}{M}\int^{R}_{0}\overline{q_{\alpha}}(\frac{\chi
r_{x}}{R})\gamma^{i}q_{\beta}(\frac{\chi
r_{x}}{R})d^{3}x\int(\mathbf{S}\times\mathbf{q})^{i}(-\frac{1}{|\mathbf{r_{x}}-\mathbf{r_{y}|}}+\frac{k}{R})|\Psi(\mathbf{y})|^{2}d^{3}y
\end{eqnarray}
where $C_{\alpha\beta}$ refers to the spin projections of the
quarks in the baryon. The transitional momentum $\mathbf{q}$ will
change into the form $-i\nabla$ in configurational space acting on
the relative position, and we then get:
\begin{eqnarray}
H_{AA\frac{1}{2}}&=&g^{2}_{s}N\int^{R}_{0}[j_{0}^{2}(\frac{\chi
r_{x}}{R})+j_{1}^{2}(\frac{\chi
r_{x}}{R})]d^{3}x\int(-\frac{1}{|\mathbf{r_{x}-\mathbf{r_{y}}}|}+\frac{k}{R})|\Psi(y)|^{2}d^{3}y
\nonumber \\&&
{}+g^{2}_{s}\frac{C_{\alpha\beta}}{M}\int^{R}_{0}\overline{q}_{\alpha}(\frac{\chi
r_{x}}{R})\gamma^{i}q_{\beta}(\frac{\chi
r_{x}}{R})d^{3}x\int\frac{[\mathbf{s}\times(\mathbf{r_{x}}-\mathbf{r_{y}})]^{i}}{|\mathbf{r_{x}}-\mathbf{r_{y}}|^{3}}|\Psi(\mathbf{y})|^{2}d^{3}y
\end{eqnarray}
where ${\bf q}$ is the exchanged momentum between the quark and
diquark, and in the configuration space of the bag, it is an
operator acting only on the relative coordinate ${\bf r}$ as
$-i\nabla_{\bf r}$ and
\begin{eqnarray}
\int
e^{i\mathbf{q}\cdot\mathbf{r}}\frac{\mathbf{q}}{|\mathbf{q}|^{2}}
\frac{d^{3}q}{(2\pi)^{3}}=\frac{i\mathbf{r}}{r^{3}}.
\end{eqnarray}
We finally obtain an integral which must be carried out
numerically if we take the form of $|\Psi(y)|^{2}$ which is
treated as $\delta$ function into our consider, we will finally
read:
\begin{eqnarray}
&H_{AA\frac{1}{2}}& = g^{2}_{s}N\int^{R}_{0}[j_{0}^{2}(\frac{\chi
r_{x}}{R})+j_{1}^{2}(\frac{\chi
r_{x}}{R})](-\frac{1}{r_{x}}+\frac{k}{R})d^{3}x \nonumber \\&&
{}+g^{2}_{s}\frac{N'}{M}\int^{R}_{0}j_{0}(\frac{\chi
r_{x}}{R})j_{1}(\frac{\chi
r_{x}}{R})<s_{1}m'_{1}s_{2}m'_{2}|[\frac{(\mathbf{\sigma}\cdot\mathbf{S})}{r_{x}^{2}}\nonumber
\\&&
{}-\frac{(\mathbf{\sigma}\cdot\mathbf{r_{x}})(\mathbf{S}\cdot\mathbf{r_{x}})}{r_{x}^{4}}]|s_{1}m_{1}s_{2}m_{2}>d^{3}x
\end{eqnarray}

For the baryon of spin 3/2, which is composed of the light quark
of spin 1/2 and an axial vector diquark of spin 1, the interaction
energy is
\begin{eqnarray}
&H_{AA\frac{3}{2}}& = g^{2}_{s}N\int^{R}_{0}[j_{0}^{2}(\frac{\chi
r_{x}}{R})+j_{1}^{2}(\frac{\chi r_{x}}{R})]\frac{2}{r_{x}}d^{3}x
\nonumber \\&&
{}+g^{2}_{s}\frac{N'}{M}\int^{R}_{0}j_{0}(\frac{\chi
r_{x}}{R})j_{1}(\frac{\chi
r_{x}}{R})<\frac{1}{2},\frac{1}{2},1,1|[\frac{(\mathbf{\sigma}\cdot\mathbf{S})}{r_{x}^{2}}\nonumber
\\&&
{}-\frac{(\mathbf{\sigma}\cdot\mathbf{r_{x}})(\mathbf{S}\cdot\mathbf{r_{x}})}{r_{x}^{4}}]|\frac{1}{2},\frac{1}{2},1,1>d^{3}x
\end{eqnarray}

For a mixing hamiltonian which originates from the transition
$S+q\rightarrow A+q$, the non-diagonal interaction element is
\begin{eqnarray}
H_{AS}&=&\int\int\overline{q}(\mathbf{x})\gamma^{\mu}q(\mathbf{x})D_{\mu\nu}\Psi^{\ast}(\mathbf{y})(i\epsilon^{\nu
l\rho\sigma}\eta'^{\ast}_{l}v'_{\rho}v_{\sigma})\Psi(\mathbf{y})
d^{3}xd^{3}y
\end{eqnarray}
After a straightforward integration, we obtain
\begin{eqnarray}
&H_{AS}& = g^{2}_{s}\frac{N''}{M}\int^{R}_{0}j_{0}(\frac{\chi
r_{x}}{R})j_{1}(\frac{\chi
r_{x}}{R})<s_{1}m'_{1}|[\frac{(\mathbf{\sigma}\cdot\mathbf{S})}{r_{x}^{2}}\nonumber
\\&&
{}-\frac{(\mathbf{\sigma}\cdot\mathbf{r_{x}})(\mathbf{S}\cdot\mathbf{r_{x}})}{r_{x}^{4}}]|s_{1}m_{1}>d^{3}x
\end{eqnarray}
It is noted that here $\eta_{\mu}$ remains as a four-vector, and
it reduces into a three-vector ${\mbox{\boldmath $\eta$}}$ which
is not an operator.

With the mixing between $|q(bc)_0>_{1/2}$ and $|q(bc)_1>_{1/2}$,
one can write the real eigenstates of $(qbc)_{1/2}$ and
$(qbc)'_{1/2}$ as
\begin{eqnarray}
|\frac{1}{2},\frac{1}{2}>=C|\frac{1}{2},\frac{1}{2},0,0>+D(-\sqrt{\frac{1}{3}}
|\frac{1}{2},\frac{1}{2},1,0>+\sqrt{\frac{2}{3}}|\frac{1}{2},-\frac{1}{2},1,1>),
\end{eqnarray}
where C and D are the coefficients to be determined. The
interaction energy is
\begin{eqnarray}
 &<\frac{1}{2},\frac{1}{2}|H|\frac{1}{2},\frac{1}{2}>&
=C^{2}<\frac{1}{2},\frac{1}{2},0,0|H|\frac{1}{2},\frac{1}{2},0,0>+D^{2}(\frac{1}{3}<\frac{1}{2},\frac{1}{2},1,0|H|\frac{1}{2},\frac{1}{2},1,0>\nonumber
\\&&
{}+\frac{2}{3}<\frac{1}{2},-\frac{1}{2},1,1|H|\frac{1}{2},-\frac{1}{2},1,1>-\frac{2\sqrt{2}}{3}<\frac{1}{2},\frac{1}{2},1,0|H|\frac{1}{2},-\frac{1}{2},1,1>)\nonumber
\\&&{}+CD(-\sqrt{\frac{1}{3}}<\frac{1}{2},\frac{1}{2},0,0|H|\frac{1}{2},\frac{1}{2},1,0>+\sqrt{\frac{2}{3}}<\frac{1}{2},\frac{1}{2},0,0|H|\frac{1}{2},-\frac{1}{2},1,1>)\nonumber \\&& {}+
CD(-\sqrt{\frac{1}{3}}<\frac{1}{2},\frac{1}{2},1,0|H|\frac{1}{2},\frac{1}{2},0,0>\nonumber
\\&&
{}+\sqrt{\frac{2}{3}}<\frac{1}{2},-\frac{1}{2},1,1|H|\frac{1}{2},\frac{1}{2},0,0>),
\end{eqnarray}
which contains both diagonal and non-diagonal interaction
elements. It is easy to set it into a matrix form as
\begin{equation}
H=\left (
\begin{array}{cc}
H_{SS} \ H_{SA} \\
H_{AS} \  H_{AA}
\end{array}
\right ),
\end{equation}
and the corresponding Schr\"odinger equation is
\begin{equation}
\left (
\begin{array}{cc}
H_{SS} \ H_{SA} \\
H_{AS} \  H_{AA}
\end{array}
\right ) \left (
\begin{array}{cc}
C \\
D
\end{array}
\right )=E \left (
\begin{array}{cc}
C \\
D
\end{array}\right ).
\end{equation}
Diagonalizing the matrix, one can solve C and D which would
determine the fraction of $|q(bc)_0>_{1/2}$ and $|q(bc)_1>_{1/2}$
in the eigenstates.

\section{ The numerical results }

In this work, we have the input parameters as

$ m_{u}=m_{d}=m_{q}\approx 0$, $m_{s}=0.279$ GeV,

$M_{cc}=3.26\;{\rm GeV},\; M_{bb}=9.79\;{\rm GeV},\;
M_{bc}=6.52\;{\rm GeV}$ \cite{Ebert}.

$\alpha_{s}$=0.23, $B= (0.145 {\rm GeV})^{4}$.

As argued above, the zero-point energy is not considered.
Minimizing the expression of $M_B$ in eq.(1) with respect to the
bag radius $R$, we obtain an equation and then determine the
R-value. Substituting the obtained R-value into the expressions,
we achieve the following table. In the table, we list our
numerical results for the baryons which contain two heavy quarks,
meanwhile the results obtained in other approaches are presented
in the table for a clear comparison.

\vspace{0.2cm}

\begin{center}
\begin{tabular}{|c||c|c|c|c|c|c|c|c|c|} \hline
$Notation$ & $content$ & $J^{p}$ & $M_{B}$ & $M_{B}$ & $M_{B}$ & $M_{B}$  & $M_{B}$ & $M_{B}$ & $M_{B}$\\
&&&(our results)&[5]&[7]&[8]&[14]&[15]&[16]\\
\hline \hline
$\Xi_{cc}$&(cc)q&$\frac{1}{2}^{+}$&3.55&3.66&3.48&3.620&3.55&3.66&3.61\\
\hline
$\Xi^{*}_{cc}$&(cc)q&$\frac{3}{2}^{+}$&3.59&3.81&3.61&3.727&3.64&3.74&3.68\\
\hline
$\Omega_{cc}$&(cc)s&$\frac{1}{2}^{+}$&3.73&3.76&3.59&3.778&3.66&3.74&3.71\\
\hline
$\Omega^{*}_{cc}$&(cc)s&$\frac{3}{2}^{+}$&3.77&3.89&3.73&3.872&3.73&3.82&3.76\\
\hline
$\Xi_{bb}$&(bb)q&$\frac{1}{2}^{+}$&10.10&10.23&10.09&10.202&-&10.34&-\\
\hline
$\Xi^{*}_{bb}$&(bb)q&$\frac{3}{2}^{+}$&10.11&10.28&10.11&10.237&-&10.37&-\\
\hline
$\Omega_{bb}$&(bb)s&$\frac{1}{2}^{+}$&10.28&10.32&10.21&10.359&-&10.37&-\\
\hline
$\Omega^{*}_{bb}$&(bb)s&$\frac{3}{2}^{+}$&10.29&10.36&10.26&10.389&-&10.40&-\\
\hline
$\Xi_{cb}$&(cb)q&$\frac{1}{2}^{+}$&6.80&6.95&6.82&6.933&-&7.04&-\\
\hline
$\Xi'_{cb}$&(cb)q&$\frac{1}{2}^{+}$&6.87&7.00&6.85&6.963&-&6.99&-\\
\hline
$\Xi^{*}_{cb}$&(cb)q&$\frac{3}{2}^{+}$&6.85&7.02&6.90&6.980&-&7.06&-\\
\hline
$\Omega_{cb}$&(cb)s&$\frac{1}{2}^{+}$&6.98&7.05&6.93&7.088&-&7.09&-\\
\hline
$\Omega'_{cb}$&(cb)s&$\frac{1}{2}^{+}$&7.05&7.09&6.97&7.116&-&7.06&-\\
\hline
$\Omega^{*}_{cb}$&(cb)s&$\frac{3}{2}^{+}$&7.02&7.11&7.00&7.130&-&7.12&-\\
\hline

\end{tabular}
\end{center}

\vspace{0.2cm}

\centerline{Table 1. The baryon spectra}

\vspace{0.2cm}

\section{ Conclusion and Discussion}

In this work, we evaluate the spectra of baryons which contains
two heavy quarks in the MIT bag model. It is an approach parallel
to the potential model which is widely adopted for studying heavy
hadrons. One can notice from the table given in the text that the
numerical results in the two approaches are consistent with each
other by the order of magnitude, but there are obvious distinction
in numbers.

The confinement in both the potential model and the bag model is
artificially introduced which may reflect the non-perturbative QCD
behavior in certain ways, but since none of them are derived from
the first principle, one cannot expect them to be  precise and it
is reasonable that they result in different numbers which are
related to the physics pictures and phenomenological parameters.
As a matter of fact, in general the parameters are obtained by
fitting data and while calculating the spectra of the lowest
states of heavy quarkonia, various potential forms and sets of
parameters can meet data. The MIT bag model is an alternative
model and this work may be complimentary to the potential model.

No matter in the potential model or the bag model, there is a
zero-point energy problem, which manifest the vacuum property of
QCD and is not calculable at present. The zero-point energy would
determine the exact positions of each baryon and its excited
states, but dies not influence the relative distances between the
states.  In the bag model, situation is a bit different, because
$E_0\propto 1/R$, when to obtain $M_B$, we differentiate $M_B$
with respect to $R$ and the minimum determines the R-value. Since
the energy of free light quark is $\sqrt {\chi^2+(mr)^2}/R$ has
the same form of $E_0$, thus its contribution can be attributed to
the little shift of the quark mass. The gaps between various
states are not affected. In the future when data of the spectra
are accumulated, one can come back to make more accurate
adjustment of all the phenomenological parameters. In this work,
we ignore the zero-point energy as argued above.

The advantages of using the bag model to evaluate the spectra are
two-folds. First the light quark obeys the relativistic Dirac
equation, secondly, one can use the QFT to evaluate the possible
mixing between $|q(bc)_0>_{1/2}$ and $|q(bc)_1>_{1/2}$ which is
impossible in the regular QM framework. In our case, when we
calculate the relativistic corrections to the interaction between
quark and diquark, we only keep the terms up to order ${\bf p}/M$,
which results in splitting of $B_{1/2}$ and $B_{3/2}$ of the same
flavor combination. Because the diquark is very heavy and $M\gg
|{\bf p}|$, this approximation would not bring up significant
changes.

As a conclusion, we evaluate the spectra of heavy baryons
containing two heavy quarks in the MIT bag model. The results are
qualitatively consistent with that obtained in the potential
model, but numerically differ by a few percents. Moreover, we
obtain a relatively large mixing between $|q(bc)_0>_{1/2}$ and
$|q(bc)_1>_{1/2}$ which will be tested in the future measurements.
Since there are no data available so far, we cannot fix a few
parameters such as the zero-point energy, and one can definitely
expect some deviations from the real values. Once the data are
accumulated in the future, we can fin-tune the model and
parameters. Definitely, the data will provide us with valuable
information about the model and parameters.

Even though the present B-factories BaBar and Belle cannot produce
such baryons, TEVATRON and LHC which will begin running in 2007,
may measure the spectra of such baryons, especially the
long-expected Next-Linear-Collider (NLC) will offer us an
idealistic place for studying such hadron spectra and our
knowledge on the hadron structure can be greatly enriched.

\vspace{0.5cm}

\noindent Acknowledgement: This work is partially supported by the
National Natural Science Foundation of China.

\end{document}